\documentclass[conference]{IEEEtran}
\IEEEoverridecommandlockouts
\usepackage{cite}
\usepackage{amsmath,amssymb,amsfonts}
\usepackage{algorithm}
\usepackage{algorithmic}
\usepackage{graphicx}
\usepackage{textcomp}
\usepackage{xcolor}
\usepackage{cite}
\usepackage{siunitx}
\usepackage{graphicx}
\usepackage{tabularx}
\usepackage{float}
\usepackage{subfigure}
\usepackage{url}
\usepackage{outlines}
\usepackage{hyperref}
\usepackage{enumitem}
\usepackage{csquotes}
\usepackage{amsmath,amssymb,amsfonts}
\usepackage{algorithmic}
\usepackage{graphicx}
\usepackage{caption}
\usepackage{textcomp}
\usepackage{xcolor}
\newenvironment{conditions}
{\par\vspace{\abovedisplayskip}\noindent\begin{tabular}{>{$}l<{$} @{${}={}$} l}}
	{\end{tabular}\par\vspace{\belowdisplayskip}}

\def\BibTeX{{\rm B\kern-.05em{\sc i\kern-.025em b}\kern-.08em
    T\kern-.1667em\lower.7ex\hbox{E}\kern-.125emX}}
\begin{document}

\title{Discovery of Perception Performance Limiting Triggering Conditions in Automated Driving\\
\thanks{The research leading to these results has received funding from the European Unions Horizon 2020 research and innovation program under the Marie Sko\l dowska-Curie grant agreement No 812.788 (MSCA-ETN SAS). This publication reflects only the authors view, exempting the European Union from any liability. Project website: http://etn-sas.eu/.}
}
\author{\IEEEauthorblockN{Ahmad Adee}
	\IEEEauthorblockA{\textit{Corporate Research,} \\
		\textit{Robert Bosch GmbH,}\\
		Renningen, Germany \\
		Ahmad.Adee@de.bosch.com}
	
	\and
	\IEEEauthorblockN{Roman Gansch}
	\IEEEauthorblockA{\textit{Corporate Research,} \\
		\textit{Robert Bosch GmbH,}\\
		Renningen, Germany \\
		Roman.Gansch@de.bosch.com}
	
	\and
	\IEEEauthorblockN{Peter Liggesmeyer}
	\IEEEauthorblockA{\textit{Software Engineering Institute} \\
		\textit{University of Kaiserslautern,}\\
		Kaiserslautern, Germany \\
		liggesmeyer@cs.uni-kl.de}
	
	\and
	\IEEEauthorblockN{Claudius Glaeser}
	\IEEEauthorblockA{\textit{Corporate Research,} \\
		\textit{Robert Bosch GmbH,}\\
		Renningen, Germany \\
		Claudius.Glaeser@de.bosch.com}

	\and
	\IEEEauthorblockN{Florian Drews}
	\IEEEauthorblockA{\textit{Corporate Research,} \\
		\textit{Robert Bosch GmbH,}\\
		Renningen, Germany \\
		Florian.Drews@de.bosch.com}
}

\maketitle

\begin{abstract}
Highly automated driving (HAD) vehicles are complex systems operating in an open context. Performance limitations originating from sensing and understanding the open context under triggering conditions may result in unsafe behavior, thus, need to be identified and modeled. This aspect of safety is also discussed in standardization activities such as ISO 21448, safety of the intended functionality (SOTIF). Although SOTIF provides a non-exhaustive list of scenario factors to identify and analyze performance limitations under triggering conditions, no concrete methodology is yet provided to identify novel triggering conditions. 

We propose a methodology to identify and model novel triggering conditions in a scene in order to assess SOTIF using Bayesian network (BN) and p-value hypothesis testing. The experts provide the initial BN structure while the conditional belief tables (CBTs) are learned using dataset. P-value hypothesis testing is used to identify the relevant subset of scenes. These scenes are then analyzed by experts who provide potential triggering conditions present in the scenes. The novel triggering conditions are modeled in the BN and retested. As a case study, we provide p-value hypothesis testing of BN of LIDAR using real world data. 
\end{abstract}


\section{Introduction}
\label{ch:introduction}
Highly automated driving (HAD) vehicles are safety critical systems deployed in an open context. Their deployment in open context may ensue unsafe and uncertain behavior originating from limitations in perceiving and understanding the open context~\cite{burton2020mind}. As open context brings in a multitude of scenarios, due consideration should be given to model all possible scenario factors that may induce performance limitations, eventually affecting the HAD vehicle performance~\cite{ISO21448 }. 

Open context also impacts a perception system's (sensor and the associated processing algorithm) capabilities causing limitations and/or uncertainties in the performance. Characterization of a perception system in terms of safety requirements or key performance indicators (KPIs) is challenging because their performance is dependent on many external scenario factors. For example, functional performance of a LIDAR based perception system may be influenced by traffic density, occlusion and/or weather conditions~\cite{adeedccl2021}.

The international organization for standardization (ISO) published the standard, ISO 21448 road vehicles safety of the intended functionality (SOTIF)~\cite{ISO21448}. SOTIF provides guidelines for the analysis of the performance limitations under the influence of potential triggering conditions that in turn can lead to hazardous behavior. 
The standard provides a scene centric scenario factor list as a starting point. Moreover, it also includes deductive e.g., cause tree analysis (CTA) and exploratory analysis e.g., system theoretic process analysis (STPA)~\cite{leveson2018stpa} as tools to model such scenario factors. ISO 21448~\cite{ISO21448} advocates the use of exploratory analysis e.g., STPA in cases where the understanding of interaction between system and its context is insufficient. While STPA may provide a good modeling representation, it cannot substantiate the interactions by quantification, to the best knowledge of the authors. In our previous publications~\cite{adeedccl2021,adeeesrel2020}, we have proposed BN as an alternate modeling tool in this regard.

The BN~\cite{pearl2014probabilistic} is a directed acyclic graph (DAG) that consists of nodes and edges. Every node is a random variable $(X_1,\dots,X_n)$. The edges represent a directed relationship between two nodes and run from the parent node $(pa)$ towards the child node $(ch)$. Together, nodes and edges represent the structure of the probabilistic network. The strength of these dependencies is governed by conditional probability distributions $\Pr(ch\mid pa)$~\cite{koller2009probabilistic}. Mathematically, the BN can be written as follows.
\begin{equation}
	\Pr (X_1,\dots,X_n) = \prod_{i}^{n} \Pr(X_i\mid pa(X_i))
\end{equation}
BN is effective in modeling uncertainty and probabilistic reasoning of a system. It exploits the dependence relationship through the local conditions in the model to perform uncertainty analysis for prediction, classification and causal inference of scenario factors. We provide a methodology to model perception's performance limitations under triggering conditions in a given scene to assess SOTIF in our previous work~\cite{adeedccl2021}. Modeling of the scene into a BN structure is solely based on expert knowledge. The resulting BN structure is fixed and assumed as the best representation of the scene for the given dataset~\cite{adeedccl2021}. However, this assumption may not always be true. Due to the open context and a general lack of knowledge about the context and system, modeling all the triggering conditions becomes a very challenging task. Identification of novel triggering conditions becomes specifically important in case of the HAD vehicles deployed in the open context. Moreover, the open context may evolve over time and new phenomena may emerge, even if the initial triggering conditions set was sufficient. For example, the initial guess of an expert about the triggering conditions for the LIDAR performance in terms of false negative \emph{(FN)} rate can be \emph{truncation}, \emph{reflection} and \emph{occlusion}~\cite{adeedccl2021}, however, it is also entirely possible that the \emph{FN} rate is also influenced by the additional triggering conditions such as \emph{traffic density}.

In this publication, we introduce a methodology to identify and model novel triggering conditions in BN using p-value hypothesis testing. The expert elicited initial BN structure ~\cite{adeedccl2021} and learned conditional belief tables (CBTs) are subjected to p-value hypothesis testing. This testing results in a small number of relevant scenes that correspond to test dataset, which are deemed inconsistent to the initial BN structure and CBTs, thus serving as a quality metric for the initial BN. Experts then analyze the scenes in order to identify novel triggering conditions. These new expert opinions (in terms of novel triggering conditions) are modeled and validated through p-value hypothesis testing on the new BN structure and the learned CBTs. Summarizing, we test the assumption of best representation of the scene through p-value hypothesis testing.
We provide the following contributions.
\begin{itemize}
	\item We introduce a methodology to model novel triggering conditions in a scene to assess SOTIF by using p-value hypothesis testing on BN structure and learned CBTs.
	\item We introduce a scene-based definition of local neighborhood and hypothesis testing.
	\item We implement the methodology on a LIDAR case study in which we use real world dataset to accept or reject novel triggering conditions.
\end{itemize}

The publication is structured as follows: sec.~\ref{ch:methodology} presents the proposed methodology. Sec.~\ref{ch:setup} briefly describes the setup used for data acquisition. Sec.~\ref{ch:Implementation} provides the application of proposed methodology on LIDAR perception. In sec.~\ref{ch:Results}, results of the implementation are evaluated. Sec.~\ref{ch:evaluation} provides limitations of the methodology while sec.~\ref{ch:RelatedWork} presents an overview of the state of the art. Finally, in sec.~\ref{ch:conclusion} we discuss conclusion and future work.
\section{Proposed Methodology}
\label{ch:methodology}
\begin{figure}
	\centering
	\includegraphics[width=1\linewidth]{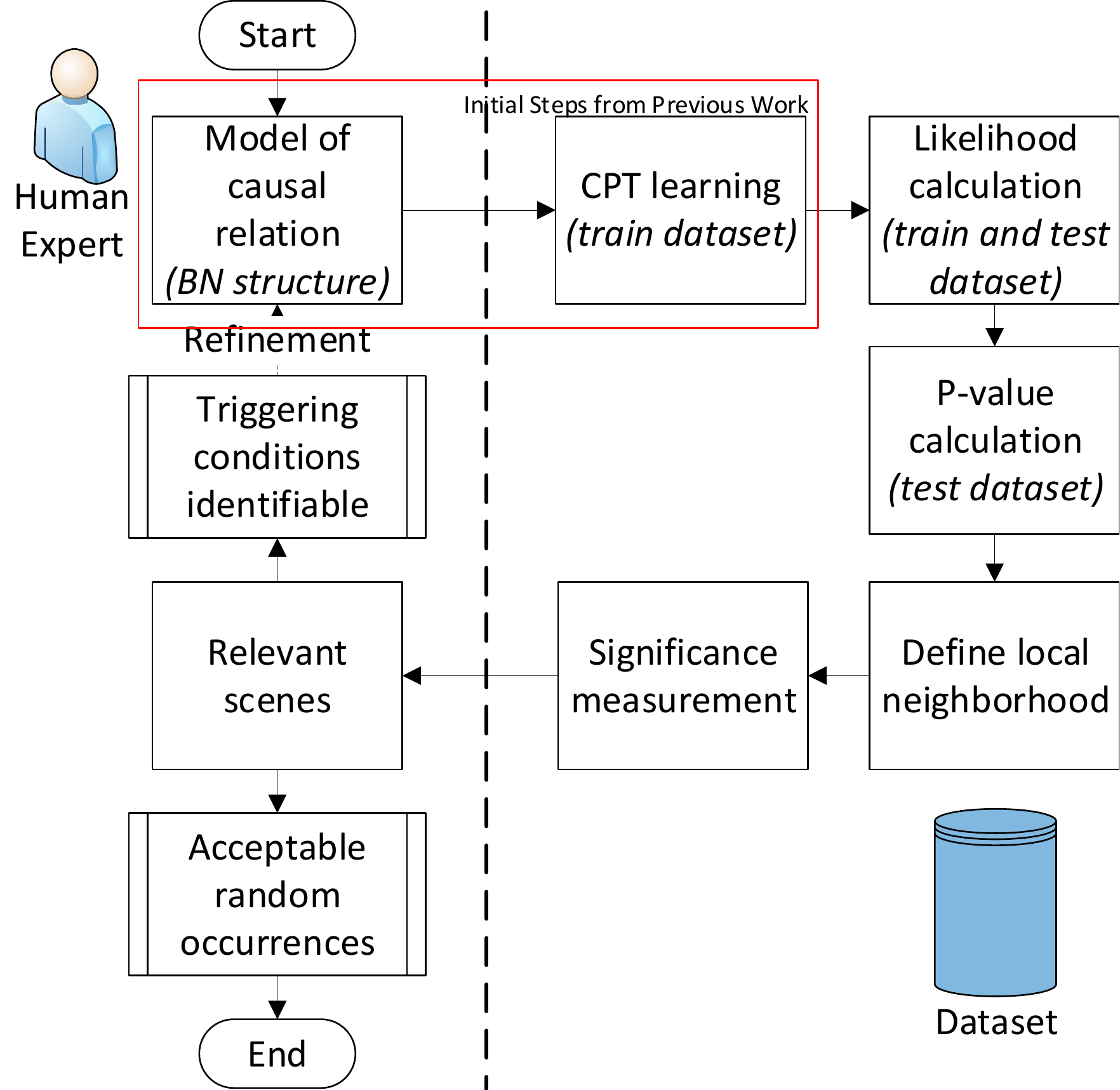}
	\caption{Flowchart describing the flow of the proposed methodology. SOTIF relevant scenario factors and expert knowledge are encoded into scene model defined by the BN structure. The shaded steps are part of previous publication~\cite{adeedccl2021}. Established CBTs (after parameter learning) are tested with p-value hypothesis and relevant scenarios are extracted, refinement steps are introduced and retested till a sufficiently accurate BN is achieved.} \label{fig:flowchart}
\end{figure}

Fig.~\ref{fig:flowchart} shows the flowchart of the methodology we adopt in this publication.
We introduce a methodology to discover novel triggering conditions and the dependence relationship in a scene with respect to the initially proposed BN. We then model, quantify and verify the potential triggering conditions.  
The experts provide the initial structure of BN while the conditional belief tables (CBTs) are learned from the training dataset. We then perform hypothesis tests on the learned BN using p-values statistics~\cite{sheskin2003handbook,mcfowland2013fast,mcfowlandautomated}, which result in subset of relevant scenes. These scenes are then analyzed by the experts, and they provide refinement strategies, accordingly. 
 Since this publication is focused on the SOTIF, we provide SOTIF relevant triggering conditions as input to model the causal relations~\cite{adeedccl2021}. However, the underlying methodology is generic and can be applied to diverse domains.
\subsection{Model of the Causal Relation}\label{subsec:methodcausalrelation}
Modeling the initial set of the casual relations is an important step of the methodology. We utilize the step described in our previous work~\cite{adeedccl2021} to model the causal relation and summarize it as follows.



SOTIF related undesired behavior (e.g., braking when it is not required and vice versa) may originate from performance limitations e.g., higher FP and FN rate~\cite{ISO21448}. Apart from FP and FN, positional error, contour matching, classification as well as regression quality can also be modeled as performance limitations.

We utilize the scenario factors from ISO 21448~\cite{ISO21448} as well as expert opinion, previous data and existing setup (constraint on data acquisition and/or data labels) to model the scene in our methodology (Fig.~\ref{fig:flowchart}). The work uses BN structure modeling for this purpose. Traditionally, the BN structure modeling is based either on the expert knowledge~\cite{flores2011incorporating} or on the learning from data (structure learning)~\cite{scanagatta2019survey}. However, in structure learning from data the number of graph candidates grow exponentially with the number of variables in the data~\cite{kabli2007chain}. Modeling causal relations is also challenging using the structure learning techniques. Due to this, we opt for the former technique in this work.

\begin{figure}
	\centering
	\includegraphics[width=0.9\linewidth]{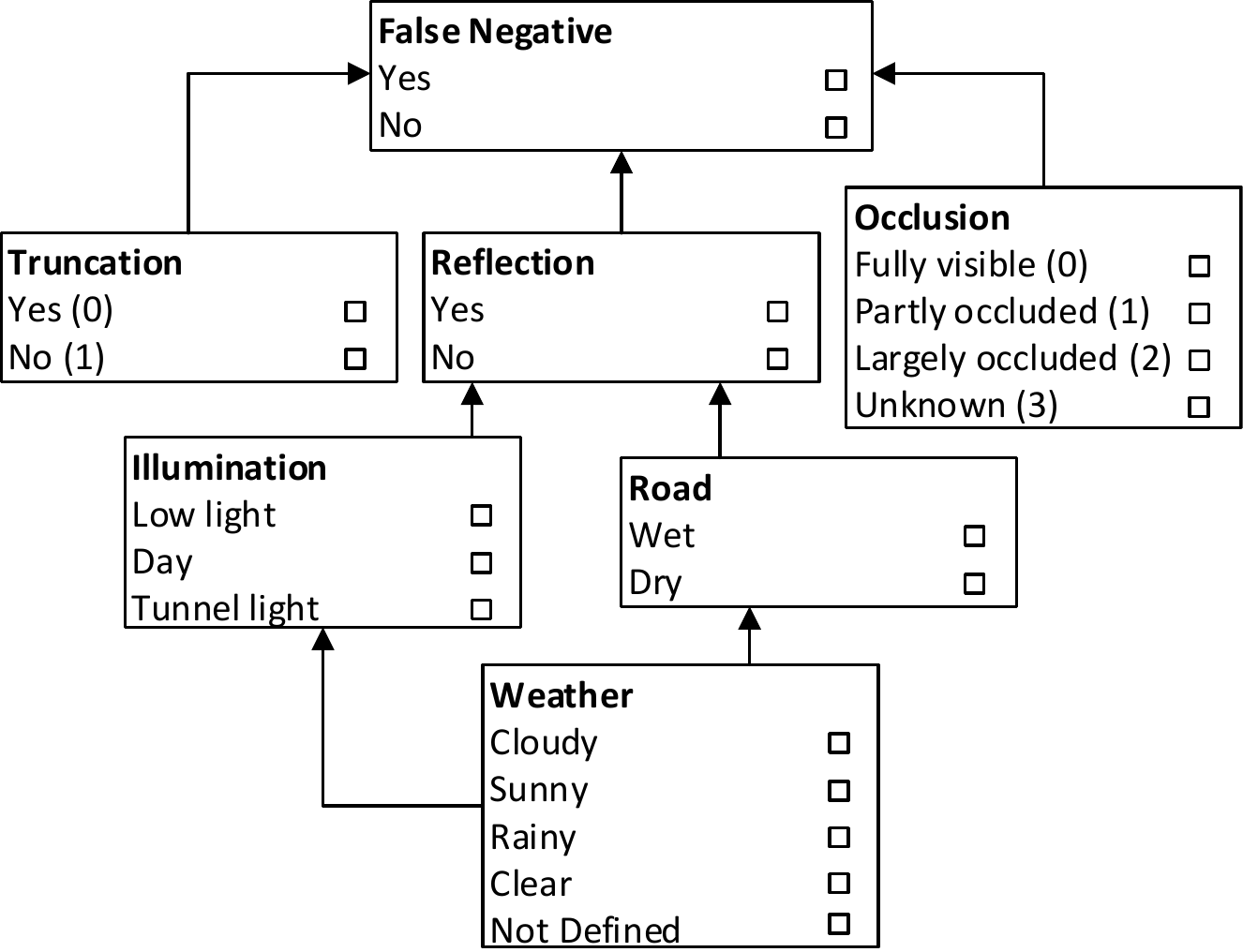}
	\caption{BN based on the SOTIF relevant scenario factors and expert knowledge describing the causal structure used in our implementation.} \label{fig:BN_Structure}
\end{figure}

Scene description which includes SOTIF relevant triggering conditions and corresponding performance limitations constitute the nodes of the BN structure. As a first step towards derivation of the structure, the experts establish hierarchical dependencies between performance limitations and triggering conditions of the scene. They then provide propositions based on these hierarchical dependencies e.g., the proposition $p1$: high occlusion may result in higher FNs. We then construct BN from these propositions e.g., the proposition $p1$ is modeled as an explicit node (Fig.~\ref{fig:BN_Structure}). 
\subsection{Parameter Learning}\label{subsec:methodparamlearn}
Dataset $\mathcal{D}$ obtained and used in this work consists of fully observed instances of the BN variables.
\begin{equation}\label{eq:dataset}
\mathcal{D} = \xi[1], \dots, \xi[M]
\end{equation}
Where $\xi[.]$ represents a data instance and $M$ represents the number of instances in $\mathcal{D}$. We use $train$ and $test$ subscripts to refer $train$ and $test$ $datasets$, respectively.

We execute ad-hoc steps to calculate any missing variables of the BN in the dataset. For example, data instances may not be labeled with FNs. 
With BN structure (Sec.~\ref{subsec:methodcausalrelation}) determined and corresponding data acquired, the CBTs can be learned. We determine the CBTs and thus the strength of the dependencies by utilizing the maximum likelihood estimator (MLE)~\cite{koller2009probabilistic}. Given a variable $X$ with parents $\textbf{U}$, we will have a parameter $\theta_{x\mid \textbf{u}}$ for each combination of $x \in Val(X)$ and $\textbf{u} \in Val(\textbf{U})$ for a CBT. The likelihood function is as follows.
\begin{equation}\label{eq:likelihood}
	\begin{aligned}
	L_X (\theta_{X \mid U}: \mathcal{D}_{train}) &= \prod_{m} \theta_{x[m]\textbf{u}[m]} \\
	&= \prod_{\textbf{u} \in Val(\textbf{U})} \prod_{x \in Val(X)} {\theta}_{x \mid \textbf{u}}^{M_{train}[\textbf{u},x]}
	\end{aligned}
\end{equation}
Here $\theta_{x \mid \textbf{u}}$ represents the parameter to be learned and $m$ represents the $m^{th}$ data instance in the dataset. 
Maximizing the likelihood function from Eq.~\ref{eq:likelihood} results in the learned parameter.
\begin{equation}\label{eq:maxlikelihood}
\theta_{x \mid \textbf{u}} = \frac{{M_{train}[\textbf{u},x]}}{{M_{train}[\textbf{u}]}}
\end{equation}
Where ${M_{train}[\textbf{u},x]}$ represents the combined occurrence of $u$ and $x$. Eq.~\ref{eq:maxlikelihood} defines the MLE. It is important to note that this learning scheme is implemented on the train dataset.
\subsection{Conditional Belief Likelihood Assignment}\label{subsec:CBLassign}
Given the BN structure (Sec.~\ref{subsec:methodcausalrelation}) and learned CBTs (Sec.~\ref{subsec:methodparamlearn}), we calculate the conditional belief likelihood.
\begin{equation}
	CBL_{x \mid \textbf{u}}^{j} = \xi[j]_{x \mid \textbf{u}} =\theta_{x \mid \textbf{u}}
\end{equation} 
In the above equation, $j \in M$, $CBL_{x \mid \textbf{u}}^{j}$ or $\xi[j]_{x \mid \textbf{u}}$ corresponds to realization of attribute $x$ given realization of its parents $\textbf{u}$ in the $jth$ data instance. For example, if $x: FN (Yes)$ given its parents nodes  $\textbf{u}:$ $Truncation$ $(Yes)$, $Reflection$ $(Yes),$ $Occlusion$ $(Largely$ $Occluded)$ corresponds to $jth$ row, then $CBL_{x \mid \textbf{u}}^{j}$ assignment corresponds to $\theta_{x \mid \textbf{u}}$. The CBL assignment is performed for both train and test datasets. 
\subsection{P-values Calculation}\label{subsec:pvalcal}
P-values calculation have been used for null hypothesis testing in BN~\cite{sheskin2003handbook,mcfowland2013fast,mcfowlandautomated}. The p-value can be defined as the probability of obtaining test results that are at least equally rare or rarer than observed results. Ranges in the p-values have been defined to handle ties in the conditional probabilities (CBLs in our case)~\cite{mcfowland2013fast,mcfowlandautomated}. The calculation of p-value ranges for the test dataset estimates relative frequency of equally rare or rarer CBL in the train dataset.
\begin{equation}
	M_{lower}^{CBL_{x \mid \textbf{u}}^{j}} = \sum_{k \in \mathcal{D}_{train}} I(CBL_{x \mid \textbf{u}}^{k} < CBL_{x \mid \textbf{u}}^{j})
\end{equation}
\begin{equation}
M_{equal}^{CBL_{x \mid \textbf{u}}^{j}} = \sum_{k \in \mathcal{D}_{train}} I(CBL_{x \mid \textbf{u}}^{k} = CBL_{x \mid \textbf{u}}^{j})
\end{equation}
Consequently, the p-value ranges can be defined as. 
\begin{equation}
\begin{aligned}
	p_{x \mid \textbf{u}}^{j} &= \lbrack p_{min}(p_{x \mid \textbf{u}}^{j}), p_{max}(p_{x \mid \textbf{u}}^{j}) \rbrack \\
							&= \left[ \frac{M_{lower}^{CBL_{x \mid \textbf{u}}^{j}}}{M_{train}+1}, \frac{M_{lower}^{CBL_{x \mid \textbf{u}}^{j}}+M_{equal}^{CBL_{x \mid \textbf{u}}^{j}}+1}{M_{train}+1} \right]
\end{aligned}
\end{equation}
Where $M_{train}$ corresponds to number of training data instances.
\subsection{Significance Calculation}
We quantify significance by statistical hypothesis testing of the calculated p-values (Sec.~\ref{subsec:pvalcal}). For a single p-value an indicator quantifying the significance at level $\alpha$ can be mathematically written as follows~\cite{mcfowland2013fast}.
\begin{equation}\label{eq:sigvar}
	n_\alpha (p) = I (p\leq \alpha)
\end{equation}
This definition can be extended to p-value ranges. Eq.~\ref{eq:sigvar} for p-value ranges can be written as follows.
\begin{equation}
n_\alpha (p_{x \mid \textbf{u}}^{j}) =
\begin{cases}
0 & \text{if $p_{min}(p_{x \mid \textbf{u}}^{j})$ $>$ $\alpha$} \\
1 & \text{if $p_{max}(p_{x \mid \textbf{u}}^{j})$ $<$ $\alpha$} \\
\frac{\alpha-p_{min}(p_{x \mid \textbf{u}}^{j})}{p_{max }(p_{x \mid \textbf{u}}^{j})-p_{min}(p_{x \mid \textbf{u}}^{j})} & \text{otherwise}
\end{cases}
\end{equation}
\subsection{Local Neighborhood Definition}
\begin{figure}
	\centering
	\includegraphics[width=1\linewidth]{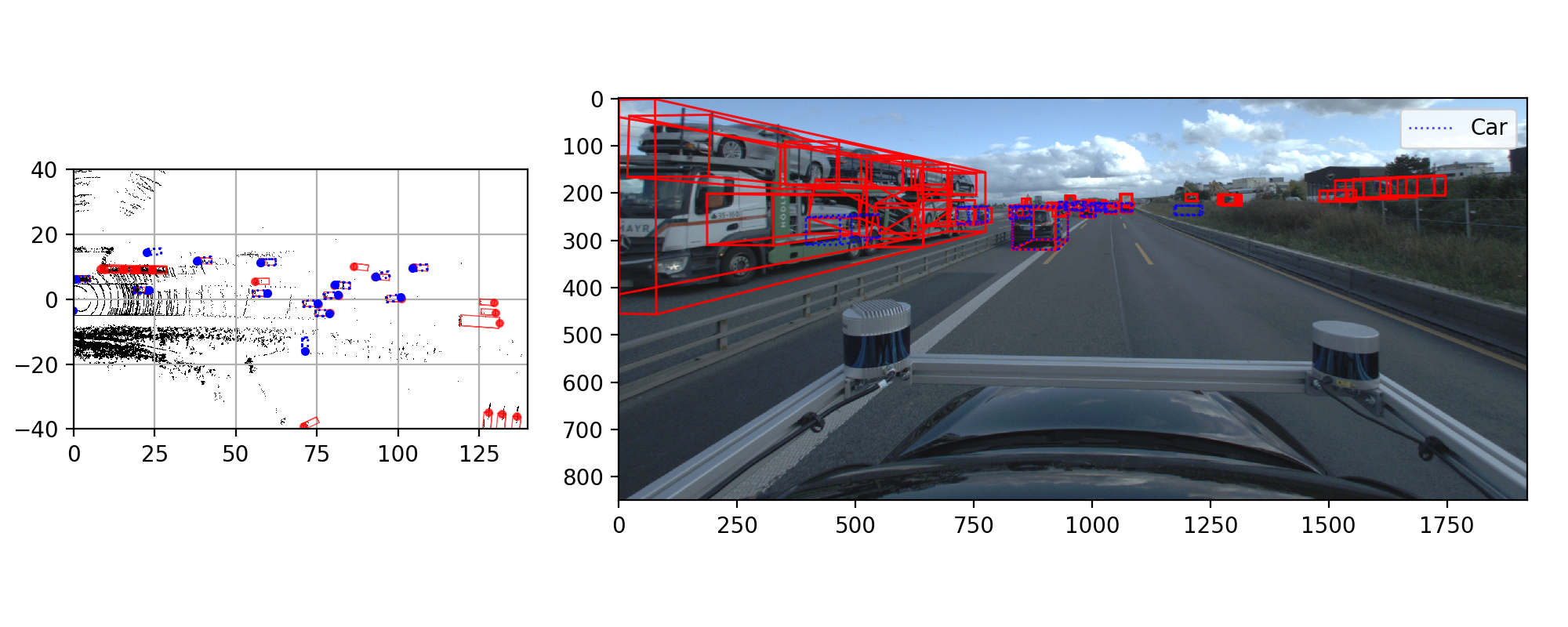}
	\caption{Scene representing the causal relations \emph{``cars loaded on a trailer"}, \emph{``ground truth labeling errors"} and \emph{``vehicle activity"}.} \label{fig:scene_ex}
\end{figure}

We believe that data instances produced by the same scene or process follow a similarity constraint, constituting a local neighborhood. Local neighborhood in the dataset can be defined by the traditional distance-based method~\cite{mcfowland2013fast} e.g., Euclidean distance for continuous and Jaccard index for categorical data instances.
However, we introduce a scene level local neighborhood by combining the scene data instances into a distinct neighborhood. An image equivalent to one such scene is shown in Fig.~\ref{fig:scene_ex}. The red annotated bounding boxes represent ground truth data instances while the blue annotated bounding boxes represent detection data instances of LIDAR datasets.
This definition of local neighborhood emphasizes that scenes in the test dataset can be as different as the significance level $\alpha$. 
\subsection{Relevant Scene Identification}\label{subsec:relsciden}
Once we establish the local neighborhood, we calculate the relevant scenes to be analyzed, by using the following mathematical notations for a local neighborhood $S$ defined by a scene.
\begin{equation}
	N_{\alpha}(S) = \sum n_\alpha (p_{x \mid \textbf{u}}^{j})
\end{equation}
\begin{equation}
N(S) = \sum I
\end{equation}
Scenes that satisfy the inequality $N_{\alpha}(S)>\alpha N(S)$ are considered relevant scenes and further analyzed. Specifically, the relevant scenes define rarer p-values in the obtained test results. 

This step can be provided for any number of nodes in the initial BN structure i.e., this step can be performed for any node/combination of nodes in the BN shown in Fig.~\ref{fig:BN_Structure}. However, in this work we confine to a single node of the initial BN within a given iteration/ implementation of the methodology.
\subsection{Relevant Scene Causal Relation}\label{subsec:relsccaurel}
Every scene identified in the previous section is subject to expert analysis. In our proposed methodology, the experts assess the scene under two probable explanations, which are discussed as follows. 
\subsubsection{Acceptable Random Occurrences}\label{subsubsec:syscaunotiden}
Experts may find scenes in which no novel triggering condition is identifiable. These scenes are considered as acceptable random occurrences. This means that no novel triggering condition can be identified by the expert through the analysis of this scene. 
\subsubsection{Novel Triggering Condition Identifiable}\label{subsubsec:occnotatran}
Experts may find scenes in which novel triggering conditions are identifiable i.e., they define relevant triggering conditions that should be taken into account in the BN structure to assess SOTIF. For example, the experts may observe traffic density as a triggering condition in the extracted scenes, that needs to be modeled in the BN. 
\subsection{Refinement}\label{subsec:refin}
Triggering conditions identified by the experts (Sec.~\ref{subsubsec:occnotatran}) are then modeled into the BN structure through the refinement step. 
Koller et al.~\cite{koller2009probabilistic} define four possible edge trails to model a variable into a BN. We only consider direct causal edge trail and confounding causal edge trail for the novel triggering conditions (Fig.~\ref{fig:refinement}). Apart from this, we also test the initial BN structure by removing a triggering condition.
\subsubsection{Direct Causal Edge Trail (DCET)}\label{subsubsec:dircetrail}
This is the simplest case, in which the novel triggering condition directly effects a node in the existing BN (Fig.~\ref{subfig:refdir}).
\subsubsection{Confounding Causal Edge Trail (CCET)}\label{subsubsec:conedtrail}
\begin{figure}
	\centering
	\subfigure[]{\includegraphics[width=0.32\textwidth]{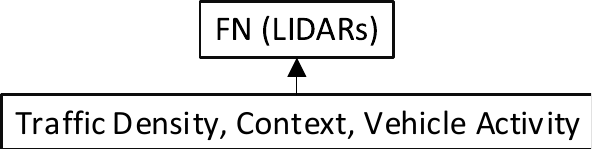}\label{subfig:refdir}} 
	\subfigure[]{\includegraphics[width=0.33\textwidth]{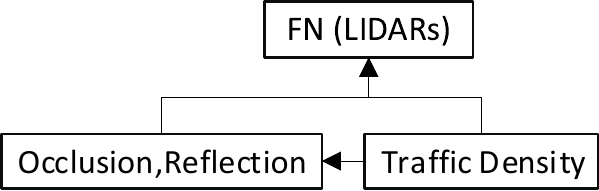}\label{subfig:refcon}}
	\caption{Refinement steps considered in this publication (a) Direct causal edge trail (DCET) (b) Confounding causal edge trail (CCET)}
	\label{fig:refinement}
\end{figure}

A confounding variable is a variable that influences both the dependent variable and independent variable, causing a spurious association (Fig.~\ref{subfig:refcon}). This occurs when the novel triggering condition controls two nodes in the existing BN simultaneously. 

Mathematically, CCET will produce a similar number of relevant scenes as DCET for a given variable in a hypothesis test, since the CCET does not change the parents of the variable. However, if the p-value hypothesis testing is performed for both independent (parent node) and dependent variable (child node) and a DCET is established for both variables, the validation results can be indicative of confounding phenomena. For example, suppose {FN} has a parent node {occlusion} while the human expert identifies {traffic density} as the novel triggering condition. In order to establish a CCET, we have to validate the DCET of {traffic density} for both {FN} and {occlusion}.
\subsubsection{Triggering Condition Removal}
This refinement step challenges the initial BN structure proposed by the experts (Sec.~\ref{subsec:methodcausalrelation}). Some nodes introduced in the initial BN structure may not validate the hypothesis test. For example, truncation, while included initially by the experts (Fig.~\ref{fig:BN_Structure}), may not be a relevant triggering condition given the dataset.
\subsection{Validation}\label{subsec:result}
Once the refinement step is taken, the BN can be tested for the relevant scene score (RSS) before and after the adjustment performed in the refinement step. This step of the implementation requires labeled data for the identified triggering conditions in order to learn the CBTs, which we consider to be made available in an iterative development approach.
The validation process includes defining a valid $proposition_{NTC}$ for a novel triggering condition from an algorithmic standpoint and final conclusion from the expert's standpoint as described in Algorithm~\ref{algo:validation}.
\begin{algorithm}
	\caption{Validation Algorithm Flow}\label{algo:validation}
	\begin{algorithmic}
		\IF{($RSS^{node}_{initial}$ $>$ $RSS^{node}_{after}$)}
		\STATE $proposition_{NTC}$ = $valid$
		\STATE $Expert$ $Conclusion$ =
		\STATE $Accepted$ $Proposition$ $\OR$ $Inconclusive$ $Evidence$
		\ELSIF {($RSS^{node}_{initial}$ $<$ $RSS^{node}_{after}$)}
		\STATE $proposition_{NTC}$ = $invalid$
		\STATE $Expert$ $Conclusion$ =
		\STATE $Rejected$ $Proposition$ $\OR$ $Inconclusive$ $Evidence$
		\ENDIF
	\end{algorithmic}
\end{algorithm}

Where $NTC$ is the novel triggering condition, $RSS^{node}_{initial}$ is the relevant scene score before the modification in the BN and $RSS^{node}_{after}$ is the relevant scene score after the modification in the BN, relative to existing BN \emph{node}. It is worth noticing that a valid proposition for a novel triggering condition $proposition_{NTC}$ may or may not be accepted by the experts. The experts may accept a proposition based on the difference between $RSS^{node}_{initial}$ and $RSS^{node}_{after}$, representatives of data, past experience and knowledge etc. Novel triggering conditions that are accepted by the experts are finally included in the modified BN. This decision criteria differs significantly from purely data driven approaches, where the decision is generally based on the results of an algorithm.
\section{Experimental Setup}
\label{ch:setup}
The experimental setup consists of two Hesai Pandar 64 and two Velodyne Ultra Puck VLP-32C LIDAR sensors installed on the roof corners of a car. The recorded data consists of different labels including bounding boxes, pose, visibility state and vehicle activity among others surrounding $\ang{360}$ of the HAD vehicle. Two separate datasets are available that correspond to detection and ground truth. Every detection and ground truth are labeled as a blue and red bounding box (Fig.~\ref{fig:scene_ex}). Most of the data was collected on different highways of Europe. However, part of the collected data also belongs to urban roads. The data consists of around twenty thousand instances. 
A deep neural network (DNN) was trained and used as the processing algorithm. 
Two experts provided their opinions on LIDAR insufficiencies, triggering conditions and limitations. 
\section{Implementation}
\label{ch:Implementation}
In this section, we demonstrate the application of our methodology on the LIDAR sensing dataset discussed in the previous section. We do not discuss Sec.~\ref{subsec:CBLassign} and Sec.~\ref{subsec:pvalcal} as they are purely mathematical calculations. Moreover, we discuss Sec.~\ref{subsec:relsciden}, Sec.~\ref{subsec:relsccaurel}, Sec.~\ref{subsec:refin} and Sec.~\ref{subsec:result} as part of the result section (Sec.~\ref{ch:Results}).
\subsection{Model of the Causal Relation}
The experts provide the list that constitutes the triggering conditions that may initiate the performance limitation of LIDAR perception system (Sec.~\ref{subsec:methodcausalrelation}). Based on the available data labels, scenario factors and experts' inputs, we conclude seven variables as SOTIF relevant performance limitation and triggering conditions. The leaf node, {FN} defines the performance limitation while the rest of the causal structure describes how the triggering conditions may impact the performance limitation (Fig.~\ref{fig:BN_Structure}). Occlusion and truncation are both defined analogously to the KITTI benchmark~\cite{Fritsch2013ITSC} and both represent the visibility state of an object.

We use the BN structure from a previous publication~\cite{adeedccl2021}. The experts provide the following initial propositions for SOTIF relevant scenario factors.
\paragraph{Proposition 1} \textit{Truncation} and \textit{occlusion} in detection may influence \textit{FN} rate.
\paragraph{Proposition 2} \textit{Weather} conditions may effect \textit{road conditions} and scene \textit{illumination}, which in turn can effect the \textit{FN} rate.
\paragraph{Proposition 3} \textit{Road condition} and scene \textit{illumination} can effect \textit{reflection} in the scene, which in turn can effect the \textit{FN} rate.\\
The resulting BN structure is shown in Fig.~\ref{fig:BN_Structure}.
\subsection{Parameter Learning}
The dataset has all the labels of our initial BN structure (Fig.~\ref{fig:BN_Structure}) except FN. We calculate FN using mean squared error (MSE) for each data instance.
\begin{equation}\label{eq:MSE}
MSE = \frac{1}{n}\sum_{i=0}^{n}(Y_i - \hat{Y_i})^2
\end{equation}
Where $n$ represents number of samples, $Y_i$ represents the ground truth and $\hat{Y_i}$ represents the detection. We execute Eq.~\ref{eq:MSE} using $x$ and $y$ values of individual detection and ground truth data instance to find a correspondence between them. All those data instance from ground truth that has no correspondence from detection using MSE are considered to be FN. 
We consider data instances with $|x|<140$ meters and $|y|<50$ meters as the DNN was trained for this range.
We perform random division of train and test datasets ($80\%$ and $20\%$) in order to perform parameter learning using Eq.~\ref{eq:maxlikelihood}. 
\subsection{Significance Calculation}
In order to perform significance calculation, we choose significance level $\alpha$ at $5\%$. We make this choice purely on the premise that p-value hypothesis testing is also performed at the same level of $\alpha$ in the state of the art~\cite{mcfowland2013fast}.
\subsection{Local Neighborhood Definition}
We define local neighborhood of test datasets based on the scenes they represent. This selection of local neighborhood equips us to identify the relevant scenes which are subject to refinement (Sec.~\ref{subsec:resultsrelsciden}) rather than identification of distance-based subsets of data instances. One such scene is shown in Fig.~\ref{fig:scene_ex}. 
\section{Results}
\label{ch:Results}
In this section, we present the results obtained by the application of our methodology. For simplicity and completeness, we adopt the following patterns in the results.
\begin{itemize}
	\item We provide the hypothesis testing and relevant scene identification for \emph{FN}, \emph{truncation}, \emph{occlusion} and \emph{reflection} nodes from the initial BN (Fig.~\ref{fig:plot_before}).
	\item We select \emph{traffic density}, \emph{vehicle activity} and \emph{context} from the relevant scene causal relation step for refinement and validation (Fig.~\ref{fig:anomaly_plots}). Moreover, we perform the refinement step for all three selected casual relations with respect to \emph{FN} only.
	\item We perform CCET refinement and corresponding validation on \emph{traffic density} with \emph{FN} and its two parent nodes i.e., \emph{reflection} and \emph{occlusion} (Fig.~\ref{fig:plot_confound}). 
	\item We perform triggering condition removal step on the \textit{truncation} node (Fig.~\ref{fig:anomaly_plots}).
\end{itemize} 

\subsection{Relevant Scene Identification}\label{subsec:resultsrelsciden}
\begin{figure}[h]
	\centering
	\includegraphics[width=0.5\linewidth]{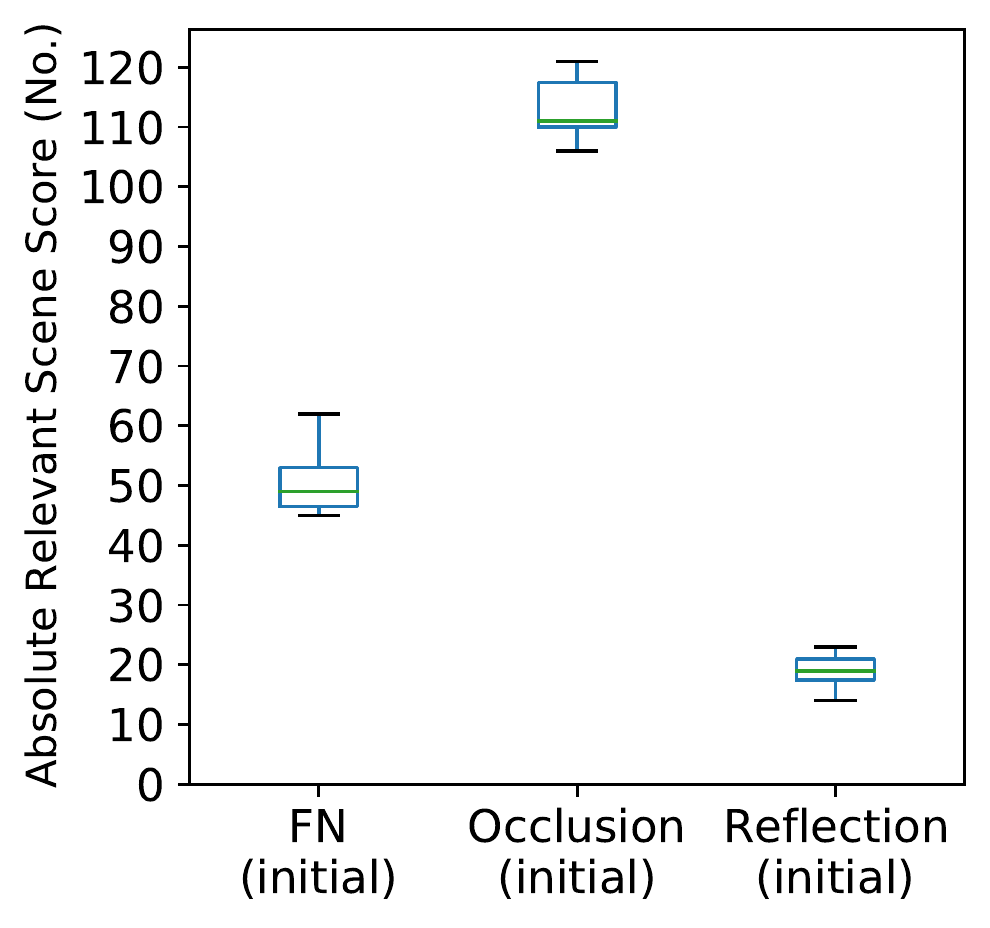}
	\caption{Relevant Scene Score (RSS) for hypothesis testing of {FN}, {occlusion} and {reflection} RSS here represents the numbers of scenes rarer at significance level $\alpha$ before any modification in the BN structure.} \label{fig:plot_before}
\end{figure}

We identify varying number of relevant scenes depending upon the randomization on the train and test dataset selection as well as the relevant nodes for which the relevant scene identification is performed. The $RSS^{node}_{initial}$ for \emph{node:} {FN}, {occlusion} and {reflection} is shown in Fig.~\ref{fig:plot_before}. There is a general decrease in the number of identified scenes (3.68\% of the test dataset scenes on average). A relatively small number of relevant scenes indicate that the test dataset is similar to train dataset.
\subsection{Relevant Scene Causal Relations}
We believe that the higher number of RSS provides the best opportunity to experts to identify relevant scene causal relations. Thus, we take the cases in which the RSS is the highest.
\begin{table}[htbp]
	\caption{Relevant scene causal relations and the respective acceptable random occurrences and identifiable novel triggering conditions cases. Identification of the triggering conditions is based on the expert opinion.}
	\begin{center}
		\begin{tabular}{|c|ccc|}
			\hline
			 Relevant Scene Causal Relations	& FN & Occlusion & Reflection\\
			\cline{1-4} 
			Acceptable Random 		& 29	  & 56	 & 10    \\
			Occurrences & & &\\
			 \hline
			Novel Triggering Condition  		& 33  & 65	 & 13	\\
						 Identifiable & & &\\
			\hline
		\end{tabular}
		\label{tab:scenediv}
	\end{center}
\end{table}
\subsubsection{Acceptable Random Occurrences}
The expert analysis of the identified relevant scenes results in 29, 56 and 10 scenes from {FN}, {occlusion} and {reflection} tests in which no novel triggering condition can be identified (Tab.~\ref{tab:scenediv}). We believe that the results do not assert that no novel triggering condition is present in these scenes. We merely believe that the experts cannot provide a probable explanation of relevancy from the novel triggering conditions standpoint.
\subsubsection{Novel Triggering Condition Identifiable}
The expert analysis of the identified relevant scenes results in 33, 65 and 13 scenes from {FN}, {occlusion} and {reflection} tests in which some novel triggering conditions can be identified (Tab.~\ref{tab:scenediv}). The identified novel triggering conditions for FN hypothesis testing are listed in Tab.~\ref{tab:scenecause}. A representative scene containing three novel triggering condition from the expert guess \emph{``cars loaded on a trailer"}, \emph{``ground truth labeling errors"} and \emph{``vehicle activity"} is shown in Fig.~\ref{fig:scene_ex}. 
\begin{table}
	\caption{Identified novel triggering conditions by FN's p-value hypothesis testing. Expert analyzes the relevant scenes identified through hypothesis testing. Triggering conditions are then provided by the experts.}
	\begin{center}
		\begin{tabular}{|c|c|}
			\hline
			Identified Novel Triggering Conditions & No. of Occurrences   \\
			\hline
			Traffic Density				& 20 \\
			\hline
			Vehicle Dimensions    		& 20 \\
			\hline
			Cars loaded on a trailer    & 5	\\
			\hline
			Ground Truth Labeling Error & 8	\\
			\hline
			Lane Discretization    		& 13	\\
			\hline
			Construction    			& 2	\\
			\hline
			Divider on Road    			& 1	\\
			\hline
			Other lane height    		& 3	\\
			\hline
			Context    					& 7	\\
			\hline
			Vehicle Activity    				& 11	\\
			\hline
		\end{tabular}
		\label{tab:scenecause}
	\end{center}
\end{table}
\subsection{Refinement}\label{subsec:refinement}
Among the potentially novel triggering conditions mentioned (Tab.~\ref{tab:scenecause}), we only discuss \emph{traffic density}, \emph{context} and \emph{vehicle activity} in detail in the refinement step. We make this selection for the following reasons.
\begin{itemize}
	\item The selected causal relations have relatively high number of occurrences.
	\item The selected causal relations have labeled data available. This selection is specific to the scope of this publication. Data labels must be made available for any selection of causal relation, if required.
\end{itemize}
The experts provide the following propositions. 
\begin{itemize}
	\item \emph{Traffic density}, \emph{vehicle activity} and \emph{context} may effect the defined performance limitation i.e., \emph{FN} (Fig.\ref{subfig:refdir}).
	\item \emph{Traffic density} may effect \emph{FN}, \emph{occlusion} and \emph{reflection}. This proposition specifically focuses on the CCET case (Fig.\ref{subfig:refcon}).
	\item \emph{Truncation} may not have any effect on the defined performance limitation i.e., \emph{FN}.
\end{itemize}
By combining the results of the first two propositions and establishing DCET, the expert can establish CCET e.g., DCET of FN and occlusion with the traffic density may result in a CCET.
We implement DCET and CCET (Sec.~\ref{subsubsec:dircetrail}) for traffic density only. The propositions can be supported by the intuition of the causal relations proposed by the experts i.e., apart from the SOTIF related measure (FN), vehicle activity and context of driving are not the cause or effect of any other node in our BN. 
\subsection{Validation}
We provide validation of expert analysis by using Algorithm~\ref{algo:validation}.
\begin{figure}
	\centering
	\subfigure[]{\includegraphics[width=0.2415\textwidth]{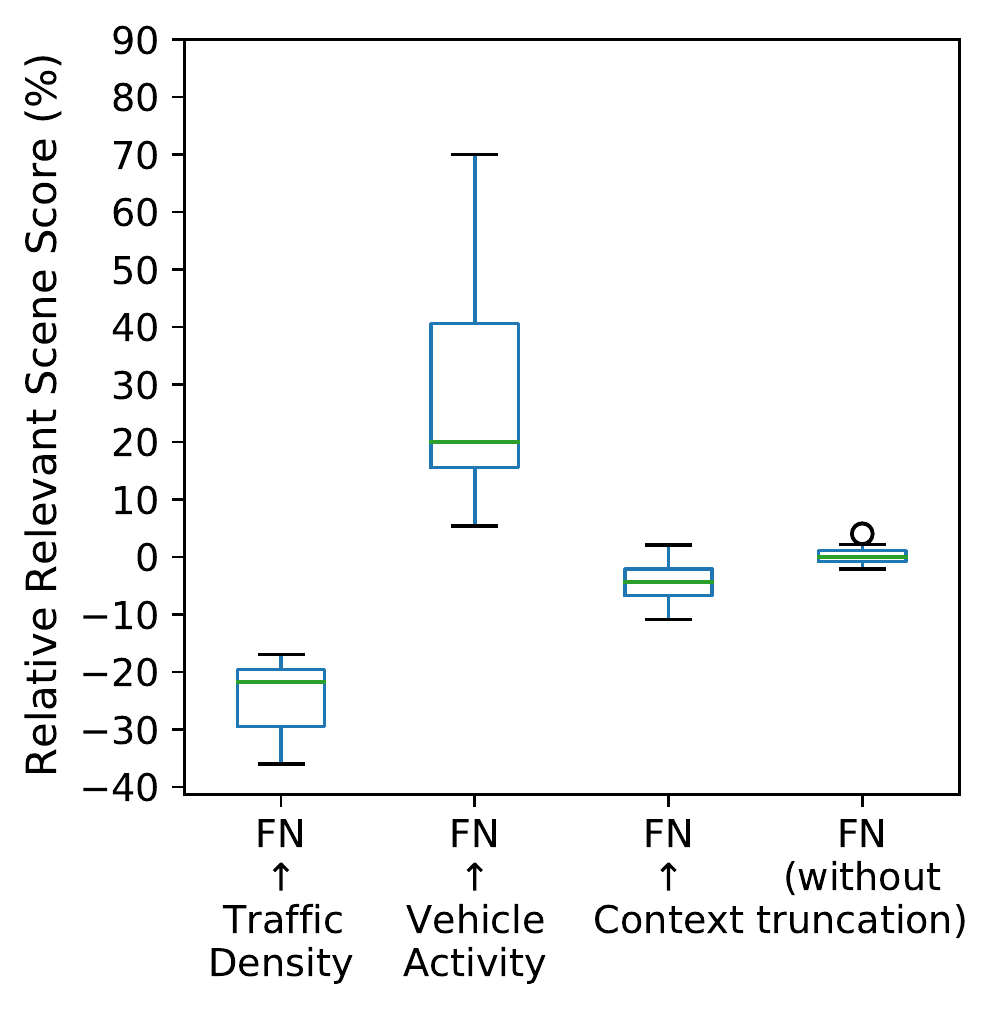}\label{fig:anomaly_plots}} 
	\subfigure[]{\includegraphics[width=0.24\textwidth]{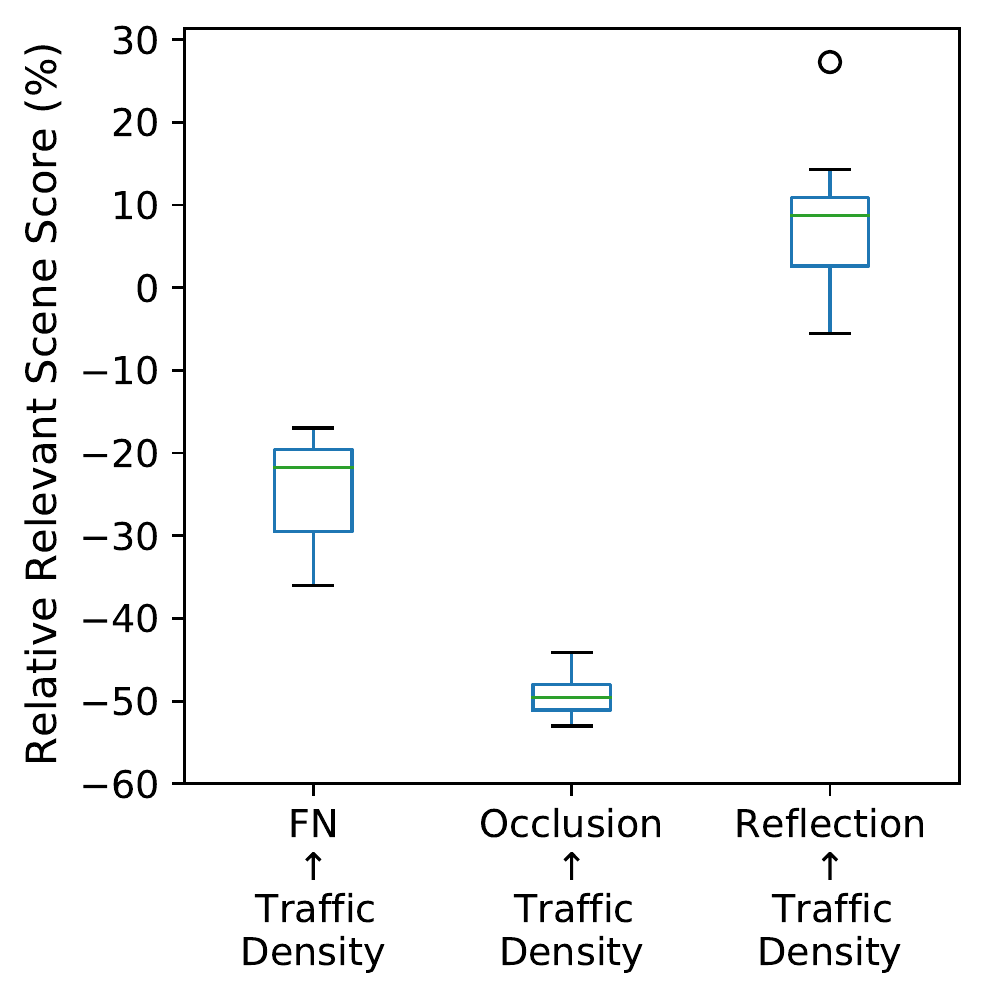}\label{fig:plot_confound}}
	\caption{\textbf{(a)} Relative relevant scene score of refinement of DCET for FN using traffic density, vehicle activity, context and without truncation as identified novel triggering conditions. Lower value after the DCET refinement indicates a more suitable BN structure than the structure considered before. \textbf{(b)} Relative relevant scene score of refinement of traffic density as DCET of FN, occlusion and reflection. The circles in the plot refer to the outlier relative RSS among the iterations.}
	\label{fig:results2}
\end{figure}
\subsubsection{Traffic Density}
We define 4 states of the traffic density; from \emph{``low''} to \emph{``very high''}. Traffic density as a triggering condition may not be very intuitive for a data instance, however, it defines the class of scenes the data instance belongs to. The validation results of the DCET (Sec.~\ref{subsubsec:dircetrail}) are shown in Fig.~\ref{fig:anomaly_plots} and Fig.~\ref{fig:plot_confound}. The following validation results can be extracted using Algorithm.~\ref{algo:validation}.
\begin{equation}\label{eq:}
\text{$proposition_{TD\to FN}$} = {valid} \impliedby \text{$RSS^{FN}_{initial}$ $>$ $RSS^{FN}_{after}$} 
\end{equation}
\begin{equation}\label{eq:}
\text{$proposition_{TD\to Occ}$} = {valid} \impliedby \text{$RSS^{Occ}_{initial}$ $>$ $RSS^{Occ}_{after}$} 
\end{equation}
\begin{equation}\label{eq:}
\text{$proposition_{TD\to Ref}$} = {invalid} \impliedby \text{$RSS^{Ref}_{initial}$ $<$ $RSS^{Ref}_{after}$} 
\end{equation}
where:
\begin{conditions}
	TD	&  traffic density \\
	Occ	&  occlusion \\   
	Ref &  reflection
\end{conditions}
The experts draw the following conclusions.
\begin{itemize}
	\item Traffic density impacts the FN rate and occlusion. 
	\item CCET trail is the most suitable construction for traffic density with occlusion and reflection.
	\item Traffic density is not a triggering condition for reflection.
\end{itemize}
\subsubsection{Context}
Two driving contexts are available in the data.
\begin{itemize}
	\item Highway: Represented by 199645 points
	\item Urban: Represented by 863 points
\end{itemize}
The validation results of the DCET from context to FN is shown in Fig.~\ref{fig:anomaly_plots}. We observe a general decrease in the relevant scenes after the adjustment in the BN structure. Algorithm~\ref{algo:validation} provides the validity of the proposition. 
\begin{equation}
\text{$proposition_{Con\to FN}$} = {valid} \impliedby \text{$RSS^{FN}_{initial}$ $>$ $RSS^{FN}_{after}$} 
\end{equation}
where:
\begin{conditions}
	Con	&  context 
\end{conditions}
However, the decrease in the RSS is not substantial, primarily due to lack of representation of the urban context in the data from the experts' standpoint. The experts draw the following conclusions.
\begin{itemize}
	\item Context of driving may impact FN rate, however, more data is required to substantiate this claim.
\end{itemize}
\subsubsection{Vehicle Activity}
Vehicle activity consists of \emph{``parked''}, \emph{``stopped''}, \emph{``moving''} and \emph{``other''} states.
The validation results of the DCET for vehicle activity as a triggering condition is shown in (Fig.~\ref{fig:anomaly_plots}). We observe an increase in the RSS after the adjustment in the BN structure. Algorithm~\ref{algo:validation} provides the validity of the proposition. 
\begin{equation}
\text{$proposition_{VA\to FN}$} = {invalid}\impliedby \text{$RSS^{FN}_{initial}$ $<$ $RSS^{FN}_{after}$} 
\end{equation}
where:
\begin{conditions}
	VA	&  vehicle activity 
\end{conditions}
The experts draw the following conclusions.
\begin{itemize}
	\item The proposition is not valid. Vehicle activity cannot be taken as a triggering condition for FN at this point.
\end{itemize}
\subsubsection{Truncation Removal}
The validation results of the DCET for truncation removal as a triggering condition is shown in (Fig.~\ref{fig:anomaly_plots}). We observe a slight increase in the RSS when truncation is not taken as a triggering condition.
Algorithm~\ref{algo:validation} provides the validity of the proposition. 
\begin{equation}
\text{$proposition_{Trun\to FN}$} = {valid} \impliedby \text{$RSS^{FN}_{initial}$ $<$ $RSS^{FN}_{after}$}
\end{equation}
where:
\begin{conditions}
	Trun	&  truncation
\end{conditions}
The experts draw the following conclusions.
\begin{itemize}
	\item Truncation removal has inconclusive evidence. More data is required to substantiate further claim.
\end{itemize}

Tab.~\ref{tab:summary} summarizes the results of our implementation. The results validated and analyzed by the experts to provide final conclusions using Algorithm~\ref{algo:validation}.
\begin{table}[h!]
	\caption{Summary of the results produced by the implementation of the methodology. Different initial nodes, systematic factors and refinements are considered in the implementation. $Abbreviations: {{Al.-Algorithm, Pro_{NTC}-proposition_{NTC}}}$}
	\begin{center}
		\begin{tabular}{|c|c|c|c|c|}
			\hline
			\textbf{Initial} & \textbf{Relative} & \textbf{Novel} & \textbf{${Pro_{NTC}}$} & \textbf{Expert}\\
			\textbf{BN} & \textbf{RSS $(\%)$} & \textbf{Triggering} & \textbf{(Al.~\ref{algo:validation})}  & \textbf{Conclusion}\\
			\textbf{Node} & & \textbf{Condition} & &\textbf{(Al.~\ref{algo:validation})}\\
			\cline{1-5} 
			FN		   & 29.02  & Vehicle	  		  & Invalid	& Rejected\\
					   &		& Activity	  		  &			& Proposition\\
			\hline
			FN		   & -4.20 	& Context	  		  &	Valid	& Inconclusive\\
					   &		&			  		  &			& Evidence\\
			\hline
			FN		   & -24.52	& Traffic	  		  & Valid	& Accepted\\
					   &		& Density	  		  &			& Proposition\\
			\hline
			Occlusion  & -49.50 & Traffic	  		  & Valid	& Accepted\\
					   &		& Density	  		  &			& Proposition\\
			\hline
			Reflection & 7.82  & Traffic	  		  & Invalid	& Rejected\\
					   &		& Density	  		  &			& Proposition\\
			\hline
			FN		   & 0.45  & \textit{Without}	  &	Invalid	& Inconclusive\\
					   &		& \textit{Truncation} &			& Evidence\\
			\hline
		\end{tabular}
		\label{tab:summary}
	\end{center}
\end{table}

\begin{figure}[h]
	\centering
	\includegraphics[width=1\linewidth]{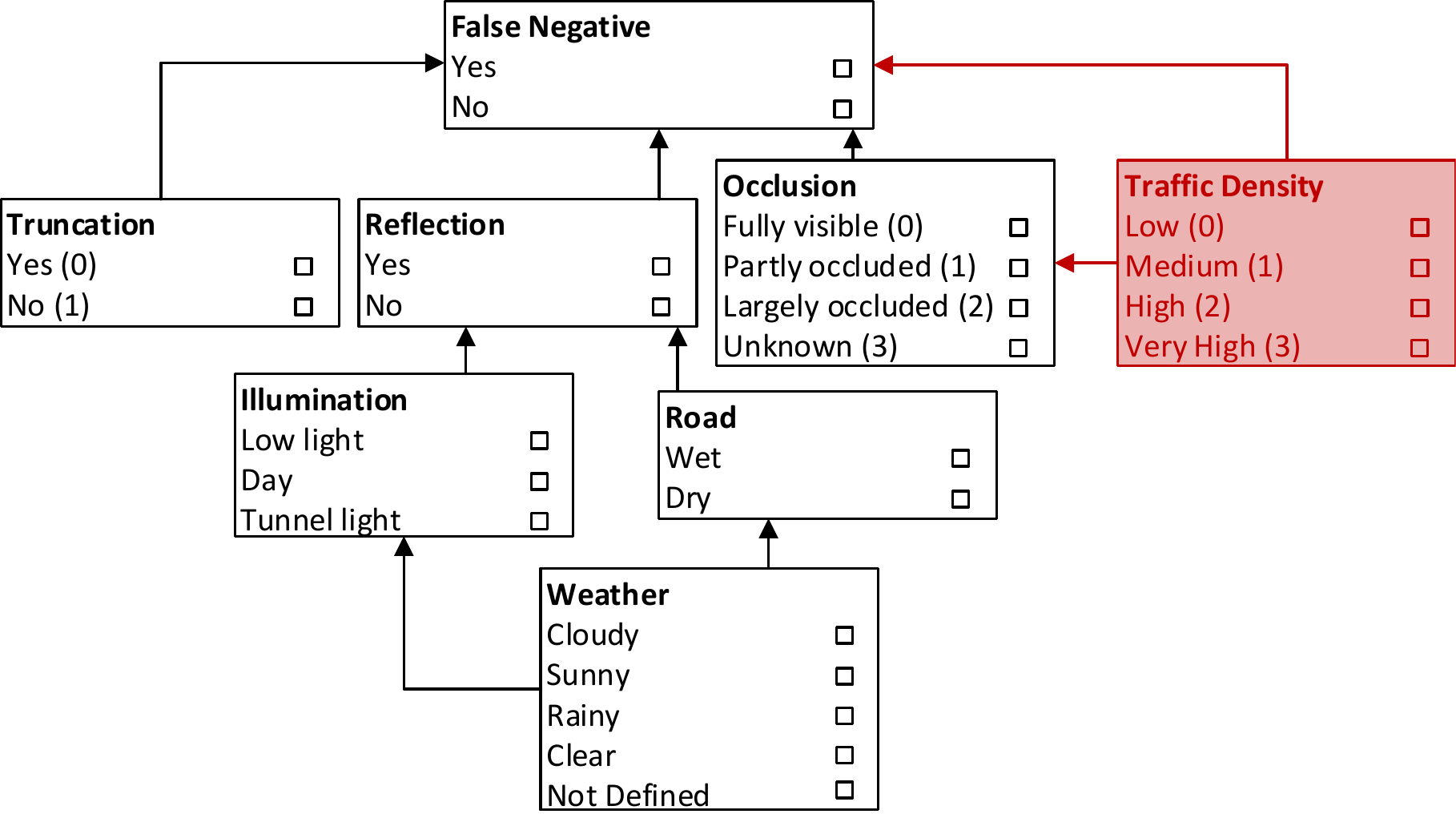}
	\caption{BN structure from Fig.~\ref{fig:BN_Structure} updated with novel triggering condition \emph{traffic density} as SOTIF relevant scenario factor for false negative and occlusion.} \label{fig:BN_Structure_update}
\end{figure}

The resulting BN structure is shown in Fig.~\ref{fig:BN_Structure_update}. We make adjustments in the BN structure by adding traffic density as CCET for FN and occlusion. With more data, the decision about driving context and truncation can also be made. In this way, novel triggering conditions can be identified, refinements can be generated and validation of those refinements can be performed in iterations to acquire more knowledge and perform a robust SOTIF analysis.

\section{Limitation of the Methodology}\label{ch:evaluation}
We perform validation of our methodology by introducing prescribed nodes in the BN structure and calculating the new RSS. We observe a substantial decrease in the RSS after modeling traffic density as a triggering condition. However, finding all the underlying triggering conditions is a challenging task. The algorithm presented in this work substantiates the triggering conditions of rare events in its true sense. We also presume that there are certain assumptions taken that are worth discussing to understand the limitation of the methodology.
\subsection{Training and Test Data}
The most important assumptions we take are related to train and test datasets. 
\begin{itemize}
	\item Train dataset does not contain any scenes that belong to rare scene family.
	\item Test dataset contains scenes that belong to rare scene family.
\end{itemize}
However, this may not always be the case. Generally, test datasets are segregated from the training datasets and they are highly correlated. We believe that randomizing the selection of datasets (e.g., Monte Carlo) can help us optimize the best solutions~\cite{ji2015monte}.
\subsection{Resemblance to Structure Learning}
A potential objection to the methodology presented can be its resemblance with the structure learning technique. Though our methodology enhances the structure by identifying novel triggering conditions, yet our methodology has the following distinct features.
\begin{itemize}
	\item We believe that a human expert plays a key role in the construction of safety analysis models. This is reflected in our methodology and is a missing feature in any structure learning technique.
	\item A human expert may identify potential triggering conditions that may not be the part of labeled data. This implies that the new BN structure may acquire causal relation not present in the initial dataset (after performing new labels). Structure learning is only confined to what is available in the form of data.
\end{itemize}
\subsection{Availability of the Labeled Data}
Availability of labeled data is very important for the proposed methodology. It may happen that the data is not available for some of the prescribed scene causal relation. For example, labels are not available for ``cars loaded on a trailer'', ``ground truth labeling error'', ``construction activity'' and ``other lane height''. We believe that the unavailability of labeled data can be addressed by the following methods.
\paragraph{Labeling Automation} Manual data labeling is a labor intensive and expensive task. However, part of the labeling process can be automated.
\paragraph{Label Ranks} If limited resources hinder the labeling process, a ranking of labels based on some structured method e.g. Phenomena Identification and Ranking Table (PIRT)~\cite{singh2019phenomena} can be used.
\subsection{Argumentation on Completeness}
Providing argumentation on the completeness of the safety model produced by our methodology to assess SOTIF is still a challenging task, despite the methods itself help in knowledge acquisition process (and thus improving completeness concerns). Even after multiple iterations of restructuring the BN, the resulting BN might not be complete. Though a conventional solution to the problem can be an expert conclusion and a sufficiency criterion defined on the RSS benchmarking, the resulting BN may not be a robust representation. Consider the following hypothetical scenario.

After restructuring the BN with the inclusion of {traffic density} as novel triggering condition, the expert conclusion and RSS satisfy the benchmark. However, it is possible that {traffic density} may not be a confounding variable and its effect on FN may be governed by a third missing variable. Such challenges can be partially solved by understanding the intuition of causal relations about real-world phenomena.


\subsection{Randomness and Lack of Knowledge Decoupling}
A general conception in the hypothesis testing is the acceptance of randomness of results to a certain significance level $\alpha$. Any value below or above the $\alpha$ (depending upon which tail of the distribution is being tested) results in unacceptable randomness and rejection of the hypothesis. In this work, we go further and instead of rejecting the hypothesis, model it with novel triggering condition. In its essence, this step corresponds to modeling lack of knowledge concepts~\cite{gansch2020}. The decoupling between randomness and lack of knowledge at some significance level $\alpha$ works with the underlying assumption i.e., any scene relation $N_\alpha (S)$ $>$ $\alpha N(S)$ is due to some triggering condition. However, it is also possible that the rarer scene occurrence is purely governed by randomness in the data. We attempt to solve this problem by allowing experts to define scenes as random occurrences (Sec.~\ref{subsubsec:syscaunotiden}).

\section{Related Work} \label{ch:RelatedWork}
Recent research indicates an ever-growing interest in SOTIF and scenario-based safety of HAD vehicles as a topic~\cite{riedmaier2020survey}. However, to the best of the authors' knowledge, existing approaches do not contribute to the knowledge acquisition process of identification, modeling, quantification and validation of novel SOTIF relevant scenario factors. Formalization of the reliability-based validation of the environment perception for safe automated driving~\cite{berk2020}, probabilistic framework for incrementally bounding the residual risk associated with autonomous drivers and its quantification~\cite{Schwalb2019AnalysisOS} and integrated method for safety assessment of automated driving functions~\cite{Kramer2020} are some of the salient literature studies in this regard. Berk et al.~\cite{berk2020} emphasizes on the failure rates of perception as well as quantification of false negative (FN) and false positive (FP) as uncertainties. Edward Schwalb~\cite{Schwalb2019AnalysisOS} focuses on continuous monitoring of SOTIF for imminent hazard by autonomous driver in order to maximize the time to materialization (TTM) by appropriate selection of actions. 
Finally, two publications~\cite{Kramer2020,neurohr2021criticality} provide identification and quantification of SOTIF related hazardous scenarios by using causal chain analysis techniques. However, the work provides a more theoretical view of the problem.

In the most recent publications~\cite{zhangtest,scholtes20216}, the major focus has been on formalization of scenario-based verification and validation. Zhang et al.~\cite{zhangtest} provide a test framework that consist of test  scenarios, different types of testing and allocation of tests, generation of test cases, data collection as well as analysis and correlation of obtained results. Scholtes et al.~\cite{scholtes20216} focuses on structured modeling of urban road environment and traffic using a six-layer model approach. 

Finally, Adee et al.~\cite{adeedccl2021}, propose a novel methodology to model triggering conditions and performance limitations in a scene to assess SOTIF using BN in this regard. The experts provide the BN structure and conditional belief tables are learned using the maximum likelihood estimator. However, the publication takes the underlying assumption that expert provided BN structure are complete and provide the best representation of the scene. 

The conceptual usage of BN and p-values hypothesis testing is performed by Mc. Flowland et al.~\cite{mcfowland2013fast,mcfowlandautomated} for anomaly pattern detection. The implementation can be distinguished from our work in the following aspects.
\begin{itemize}
	\item While Mc. Flowland et al.~\cite{mcfowland2013fast,mcfowlandautomated} use a structure learning technique for BN structure, we believe the initial BN structure should be provided by the safety experts as discussed in other literature~\cite{adeedccl2021}.
	\item In our work, the local neighborhood definition is scene based rather than a more common distance-based definition. This selection changes the entire rationale of the implementation and is a salient feature of our implementation.
	\item We close the loop by expert oriented reasoning of anomalies and provision of new BN structure that better suits our world knowledge.
	\item Moreover, the overall theme of the work by Mc. Flowland et al.~\cite{mcfowland2013fast,mcfowlandautomated} is to provide an inference algorithm for pattern detection, while we focus on the SOTIF oriented triggering condition discovery.
\end{itemize}
\section{Conclusion and Future Work}
\label{ch:conclusion}
We presented a methodology to discover novel triggering conditions under the scene model to argue  safety of the intended functionality (SOTIF) analysis. The methodology encodes parameter learning for Bayesian network (BN) and p-value testing of the learned BN. 

This methodology particularly assists in the identification of novel SOTIF related triggering conditions under manageable effort. The identified novel triggering conditions are then modeled in the BN and validated through testing.
This assists the experts to establish SOTIF modification plan for the identified performance limitation under the novel triggering conditions.

We believe that the contribution we make with this publication is very valuable from the SOTIF standpoint. Analyzing thousands of scenes to identify potentially novel triggering conditions is not feasible. Our approach curtails the number of scenes to a very small number (around \textbf{3\%} of the total in test dataset), making the analysis feasible. 

In order to argue the adequacy of the approach, LIDAR performance was studied given a scene. The scene was modeled using a BN structure and parameter learning was performed using real world data to elicit conditional belief tables (CBTs). P-value testing was performed on the learned BN and relevant scenes were extracted. These scenes were then analyzed by experts to identify the triggering conditions.

We also evaluated the decrease in the number of identified scenes and observed roughly a \textbf{25\%} decrease in case of \emph{traffic density} as newly modeled triggering condition for \emph{FN} and roughly \textbf{50\%} for \emph{occlusion}. We then discussed the limitations of the methodology.

In future, we intend to explore other hypothesis testing techniques against BN models. We also intend to provide a general framework which covers novel triggering condition identification based on rare events.

\bibliographystyle{IEEEtran}
\bibliography{IEEEabrv,references}
\end{document}